\documentclass[fleqn,12pt,twoside]{article}
\usepackage{espcrc1}

\usepackage{graphicx}
\usepackage[figuresright]{rotating}

\newcommand{\AmS}{{\protect\the\textfont2
  A\kern-.1667em\lower.5ex\hbox{M}\kern-.125emS}}

\title{Bethe-Salpeter equation with cross-ladder kernel in Minkowski
and Euclidean spaces}

\author{V.A.~Karmanov\address[Lebedev]{Lebedev Physical
Institute, \\
       Leninsky prospect 53, 119991 Moscow, Russia}%
        \thanks{Supported in part by the grants 05-02-17482-a
        and 06-02-26805 of Russian Foundation for Basic Research.},
        J.~Carbonell\address[LPSC]{Laboratoire de Physique Subatomique
        et Cosmologie,\\53 avenue des Martyrs, 38026 Grenoble,
        France}
        and
        M.~Mangin-Brinet\addressmark[LPSC]}

\begin{document}

\maketitle

\begin{abstract}
{\small Some results obtained by a new method for solving the
Bethe-Salpeter equation are presented. The method is valid for any
kernel given by irreducible Feynman graphs. The Bethe-Salpeter
amplitude, both in Minkowski and in Euclidean spaces, and the
binding energy for ladder + cross-ladder kernel are found. We
calculate also the corresponding electromagnetic form factor.}
\end{abstract}
\vspace{0.5cm}

Bethe-Salpeter (BS) equation \cite{SB_PR84_51} is an important
tool for studying the relativistic  bound state problem in a field
theory framework. The BS amplitude in Minkowski space is singular,
what makes the numerical resolution of BS equation difficult.
Therefore it is usually solved in Euclidean space to find the
binding energy.    However, to describe   some observables, we
need the original BS amplitude in Minkowski space as well. The
methods proposed so far to calculate this amplitude are valid
either for the ladder kernel \cite{KW} or for a separable one
\cite{bbmst}.

We give here a brief review of results obtained   by solving BS
equation using a new method, developed in \cite{bs12}. It allows
to find the BS amplitude both in Minkowski and Euclidean spaces as
well as the light-front (LF) wave function. The BS amplitude is
written in terms of the Nakanishi integral representation
\cite{nak63}:
\begin{equation}\label{eq1}
\Phi(k,p)=\int_{-1}^1dz'\int_0^{\infty}d\gamma'
\frac{-ig(\gamma',z')}{\left[\gamma'+m^2
-\frac{1}{4}M^2-k^2-p\cdot k\; z'-i\epsilon\right]^3}
\end{equation}
and substituted in the BS equation. The resulting equation is then
projected on the LF plane, i.e., both parts of it are integrated
over the variable $k_-=k_0-k_z$. This integration eliminates
singularities of the original BS   amplitude.    An integral
equation for the weight function $g(\gamma,z)$ is derived, which
for spinless particles has the form:
\begin{equation}
\label{bsnew} \int_0^{\infty}\frac{g(\gamma',z)d\gamma'}
{\Bigl[\gamma'+\gamma +m^2-\frac{1}{4}(1-z^2)M^2\Bigr]^2}=
\int_0^{\infty}d\gamma'\int_{-1}^{1}dz'\;V(\gamma,z;\gamma',z')
g(\gamma',z')\ ,
\end{equation}
The projected kernel $V$ is obtained in \cite{bs12} in terms of
the BS interaction kernel. This derivation does not contain any approximation.

For the massless ladder exchange the function $g(\gamma,z)$ turns
into $g(\gamma,z)=\delta(\gamma) \tilde{g}(z)$ and equation
(\ref{bsnew}) is reduced to the well-known Wick-Cutkosky equation
\cite{WICK_54} for $\tilde{g}(z)$. For massive ladder exchange,
the numerical calculations by equation (\ref{bsnew}) and by the BS
equation in Euclidean space give the same binding energy
\cite{bs12}.

\begin{figure}[htbp]
\vspace{0.5mm}
\begin{center}
\includegraphics[width=8cm]{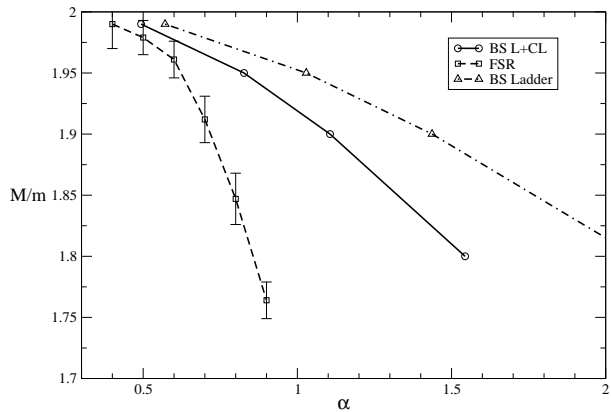}
\end{center}
\vspace{-1.1cm} \caption{Ground state mass $M$ obtained by eq.
(\protect{\ref{bsnew}}) with L and L+CL kernels, compared with the
Feynman-Schwin\-ger representation results for an exchanged mass
$\mu=0.15$.} \label{fig_B1}
\end{figure}
\vspace{-0.5cm}

Figure \ref{fig_B1} shows the ground state mass $M$ obtained for
an exchanged mass $\mu=0.15$ (in units of the constituent mass),
by eq. (\ref{bsnew}) with ladder (L) and ladder+cross-ladder
(L+CL) kernels together with Feynman-Schwin\-ger representation
results \cite{NT_PRL_96}. The latter incorporates all the higher
order cross box contributions in the kernel, but not the self
energy. The effect of CL diagrams is large and attractive, but an
even larger contribution remains to be included in the kernel, due
to the higher order terms. The L+CL calculation performed in the
framework of the light-front dynamics (LFD)  provides binding
energy very close to the BS ones.

  BS equation with L+CL kernel was also solved in Euclidean
space in \cite{bs70,cm} and  corresponding LFD   equation -- in
\cite{cmp}. Our results are in close agreement with these
references. In \cite{sfcs}, an equation obtained by projecting the
BS equation on the light-front plane was also derived.

\begin{figure}[htbp]
\vspace{-0mm}
 \centering
\includegraphics[width=7.5cm]{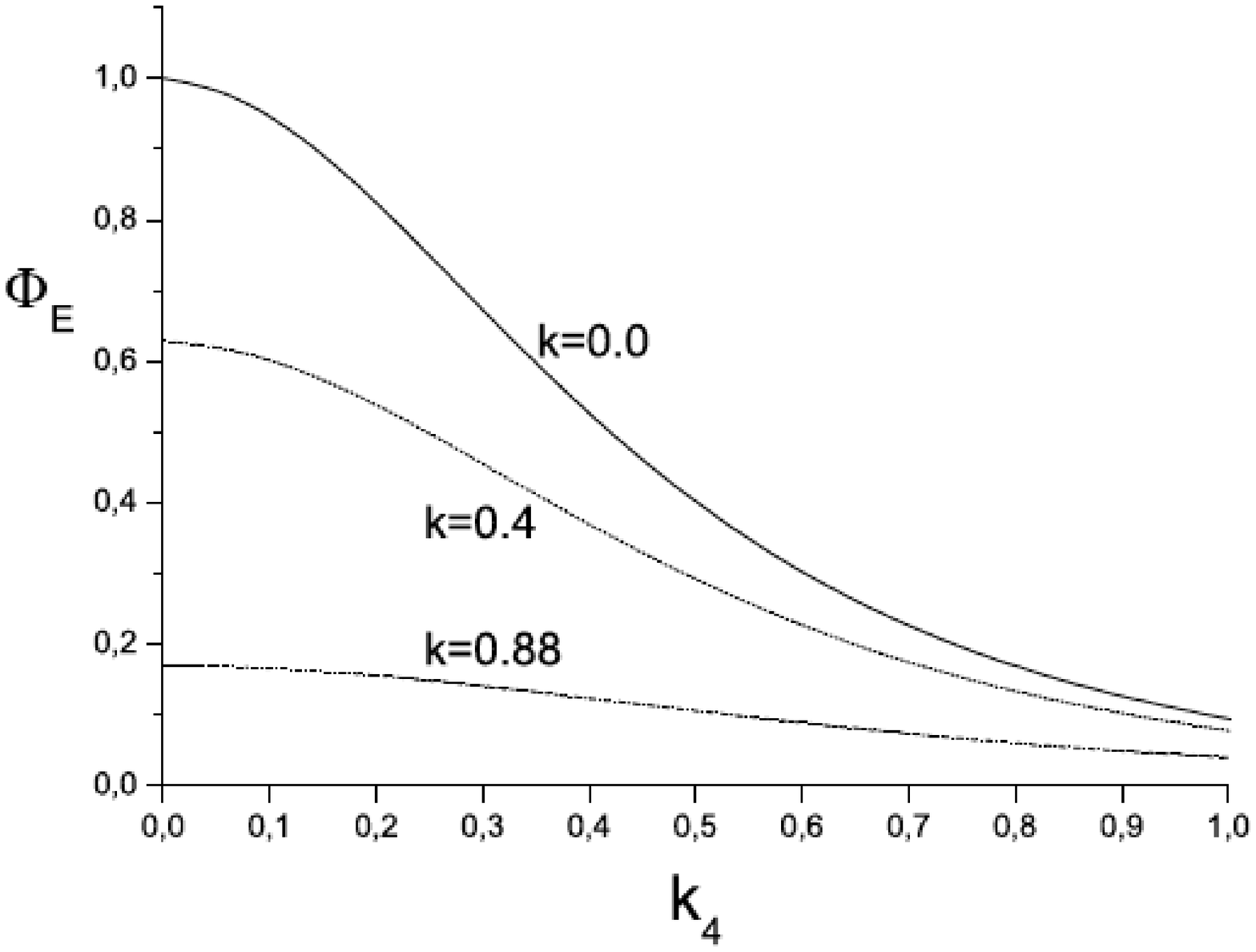}
\hspace{0.5cm}
\includegraphics[width=7.5cm]{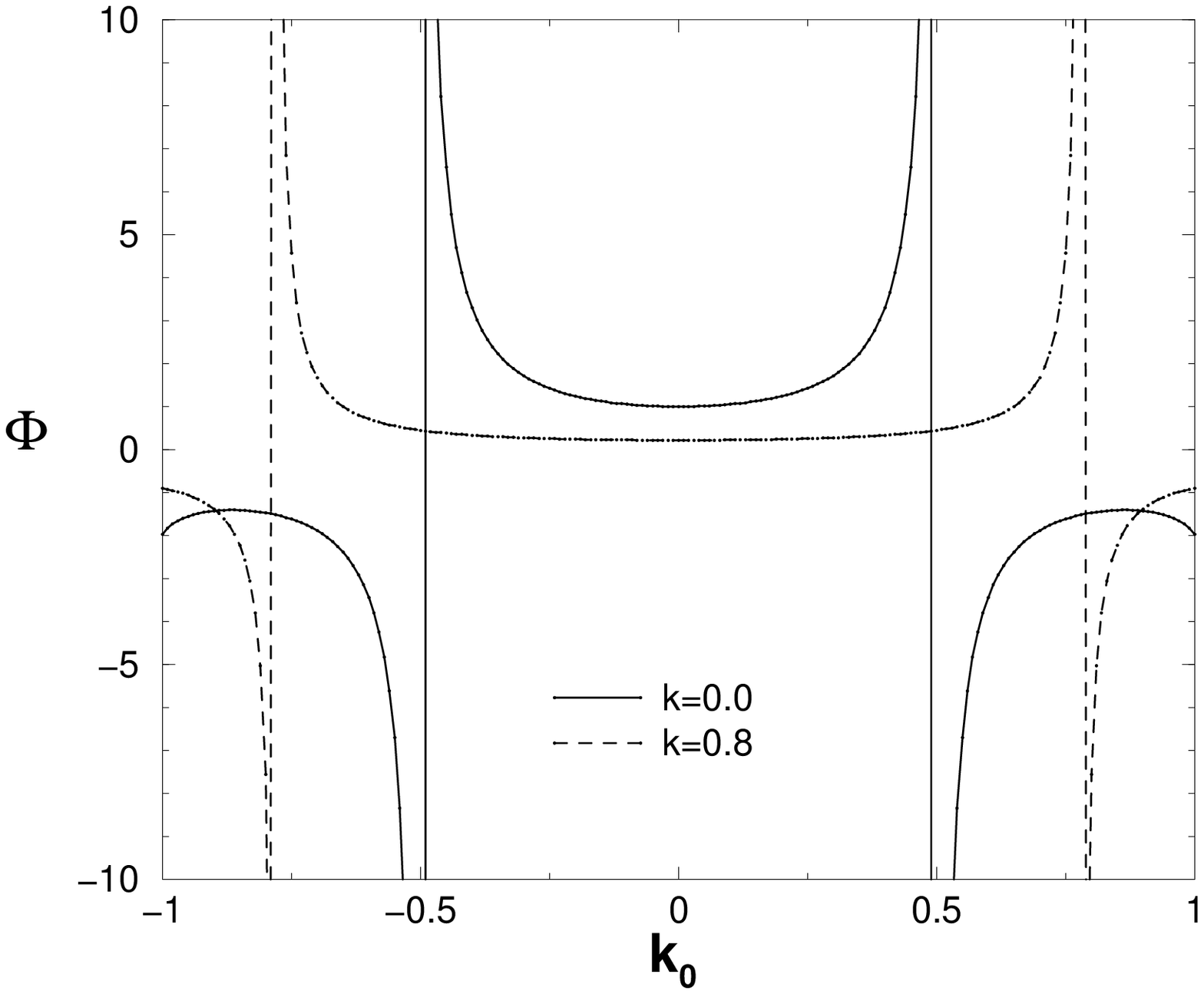}
 \vspace{-0.cm}
 \caption{Left: The Euclidean BS amplitude obtained by solving eq.
(\protect{\ref{eq1}}) with imaginary $k_0=ik_4$ (for ladder
kernel), for different values of $k$. The amplitude calculated by
direct resolution of BS equation in Euclidean space is
indistinguishable. Right: The same solution in Minkowski space,
obtained by eq. (\protect{\ref{eq1}}) for real $k_0$.}
\label{fig1}
\end{figure}

Once  $g(\gamma,z)$ is known, we can find by means of eq.
(\ref{eq1}) the BS amplitude both in Minkowski and Euclidean
spaces. The latter is obtained by substituting in this equation
$k_0=ik_4$. An example is given in Figure \ref{fig1}.   It is
worth noticing the difference between the smooth behavior of the
Euclidean amplitude and the singularities  displayed in the
Minkowski one. The amplitude in Euclidean space found by    eq.
(\ref{eq1}) for imaginary $k_0=ik_4$   coincides with one obtained
by the direct solution in Euclidean space.
We have also carried out precise calculations (accuracy better than 0.1\%) of the binding energy for L+CL kernel
  independently   by eq. (\ref{bsnew}) and in Euclidean space. The results are
presented in Table \ref{table1}. Their coincidence demonstrates
both the validity of our approach and the possibility of Wick
rotation for the CL kernel.  The latter point had not been proved
as it was the case for the ladder kernel (see a discussion in
\cite{ADT}).

\begin{table}[htb]
\vspace{-0mm}
 \caption{Comparison for BS L+CL kernel and
$\mu=0.5$, of the coupling constant $\alpha$ obtained by eq.
(\protect{\ref{bsnew}}) and by the Euclidean space calculations,
for given values of the binding energy $B$. The accuracy on the
coupling constants is better than 0.1 \% for both calculations.}
\label{table1}
\begin{tabular}{llllllll}
\hline
$B$  &0.01 &0.05 &0.1 &0.2 &0.5  &1.0  \\
\hline $\alpha$, eq.
(\protect{\ref{bsnew}}) &1.206 &1.607 &1.930 &2.416 &3.446 &4.549  \\
$\alpha$, Euclid &1.205 &1.608 &1.930 &2.417 &3.448 &4.551 \\
\hline
\end{tabular}
\end{table}
\vspace{-0.0cm}

Though the binding energy can be found from the Euclidean space
calculation, to describe many observables, we need the Minkowski
space BS amplitude. We need it even to calculate EM form factors
where, at first glance,   we could   use the Euclidean BS
amplitude, which is determined in the rest frame $\vec{p}=0$.
 Indeed, the form factors are expressed through the
Euclidean BS amplitude $\Phi_E(k_4,\vec{k};p)$ for non-zero total
momentum $\vec{p}$. To express the BS amplitude for non-zero
$\vec{p}$    via the rest frame solution
\mbox{$\Phi_E(k_4,\vec{k};p_0=M,\vec{p}=0)$}, we need to make a
boost of four-momentum $k$:
$\Phi_E(k_4,\vec{k};p)=\Phi_E(k'_4,\vec{k'};p_0=M,\vec{p}=0)$. In
Minkowski space this boost reads:
$k'_0=\frac{1}{M}(k_0p_0-\vec{k}\cdot\vec{p})$ and similarly for
$\vec{k'}$. After replacing $k_0=ik_4$ (with still real $p_0$), it
results in a complex value of boosted variables
$k'_4=\frac{1}{M}(p_0k_4+i\vec{p}\cdot \vec{k})$.
   This
requires the knowledge of the BS amplitude in the full complex
plane, and not only on the imaginary axis. The problem can be
solved in the static approximation \cite{ZT1980}, where the form
factor is calculated approximately, through the BS amplitude
obtained from the rest frame by a non-relativistic boost.
An estimation of accuracy, done in \cite{ZT1980} perturbatively,
using a decomposition in momentum transfer, shows that the
correction is not negligible and increases with momentum transfer.

\begin{figure}[htbp]
\vspace{-0mm}
\begin{center}
\includegraphics[width=8cm]{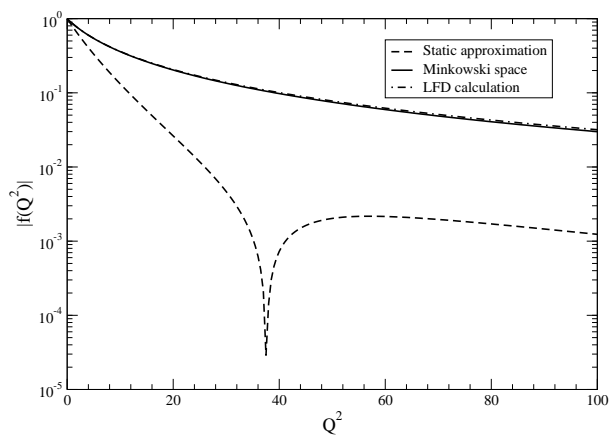}
\end{center}
\vspace{-1cm} \caption{Dashed curve: Form factor in static
approximation.  Solid curve: Exact form factor (BS in Minkowski). Dot-dashed: LFD calculation.}
\label{FF1}
\end{figure}

We calculate the form factor exactly, through the Minkowski space
BS amplitude   for L+CL,   and compare our result with the static
approximation. This comparison is shown in Figure \ref{FF1}. The
curves considerably differ from each other and the difference
increases with momentum transfer. The zero of form factor in this
model is an artifact of static approximation. The LF calculation
gives result which is almost indistinguishable from the Minkowski
space one.

We conclude that the results presented above confirm the validity of
the method \cite{bs12} and demonstrate the necessity for using
the BS amplitude in Minkowski space.

\end{document}